\documentstyle[12pt,aps]{revtex}
\newcommand{\be}{\begin{equation}}
\newcommand{\ee}{\end{equation}}
\newcommand{\bel}[1]{\be\label{#1}}
\newcommand{\re}[1]{Eq.~(\ref{#1})}
\newcommand{\mbs}[1]{\mbox{$\scriptstyle{#1}$}}
\newcommand{\ds}{\displaystyle}
\newcommand{\qq}{\mbox{$q \overline{q}$}\,\,}
\newcommand{\mub}{\overline{\mu}}
\newcommand{\psib}{\overline{\psi}}
\newcommand{\rhob}{\overline{\rho}}
\newcommand{\bm}[1]{\mbox{\boldmath${#1}$\unboldmath}}

\newcommand{\loo}{\,\raisebox{-.5ex}{$\stackrel{<}{\scriptstyle\sim}$}\,}
\newcommand{\intp}{\int\frac{{\rm d}^3 p}{(2\pi)^3}}
\newcommand{\ocnf}{n_{\bm{p}f}}
\newcommand{\ocnbf}{\overline{n}_{\bm{p}f}}
\newcommand{\dd}{\partial\hspace{-6pt}/}
\newcommand{\dD}{D\hspace{-7pt}/}
\begin{document}
\renewcommand{\thefootnote}{\arabic{footnote}}
\begin{center}
{\Large\bf Unusual bound states of quark matter\\
within the NJL model}\\[5mm]
{\bf I.N.~Mishustin$^{\,1,2}$,
L.M.~Satarov$^{\,1,3}$, H.~St\"ocker$^{\,3}$, and
W.~Greiner$^{\,3}$}
\end{center}
\begin{tabbing}
\hspace*{1.5cm}\=${}^1$\,\={\it The Kurchatov~Institute,
Russian Research Centre, 123182~Moscow,~\mbox{Russia}}\\
\>${}^2$\>{\it The Niels~Bohr~Institute,
DK--2100~Copenhagen {\O},~\mbox{Denmark}}\\
\>${}^3$\>{\it Institut~f\"{u}r~Theoretische~Physik,
J.W.~Goethe~Universit\"{a}t,}\\
\>\>{\it D--60054~Frankfurt~am~Main,~\mbox{Germany}}
\end{tabbing}

\begin{abstract}
Properties of dense quark matter in and out of chemical equilibrium are
studied within the SU(3) Nambu--Jona-Lasinio model. In addition to the
\mbox{4--fermion} scalar and vector terms the model includes also the
6--fermion flavour mixing interaction. First we study a novel form of
deconfined matter, meso-matter, which is composed of equal number of
quarks and antiquarks. It can be thought of as a strongly compressed
meson gas where mesons are melted into their elementary constituents,
quarks and antiquarks. Strongly bound states in this quark--antiquark
matter are predicted for all flavour combinations of \qq pairs. The
maximum binding energy reaches up to 180 MeV per \qq pair for mixtures
with about 70\% of strange ($s\bar{s}$) pairs. Equilibrated
baryon--rich  quark matter with various flavour compositions is also
studied. In this case only shallow bound states appear in systems with
a significant admixture (about 40\%) of strange quarks (strangelets).
Their binding energies are quite sensitive to the relative strengths of
scalar and vector interactions. The common property of all these bound
states is that they appear at high particle densities when the chiral
symmetry is nearly restored. Thermal properties of meso-matter as well
as chemically equilibrated strange quark matter are also investigated.
Possible decay modes of these bound states are discussed.

\end{abstract}

\baselineskip 24pt

\section{Introduction}
The main goal of present and future experiments with ultrarelativistic
heavy ions is to produce and study in the laboratory a new state of
strongly interacting matter, the Quark-Gluon Plasma (QGP). In achieving
this goal one is facing two major problems. First, the phase structure
of QCD is not fully understood yet. Second, the matter evolution in the
course of an ultrarelativistic heavy--ion collision may be out of
thermodynamical equilibrium.

Most calculations of the QCD phase diagram are made under the
assumption of thermal and chemical equilibrium. The QCD lattice
calculations at zero baryon chemical potential reveal a second order
phase transition or a rapid crossover at temperatures around
140--160~MeV. Recent calculations based on different effective models
\cite{Moc,Raj,Hal,Car} show the possibility of a first order phase
transition  at finite baryon densities and moderate temperatures. The
predicted phase diagram in the $T-\mu$ plane contains a first order
transition line terminating at a critical point $(T_c, \mu_c)$ with
$T_c\simeq$ 120 MeV and a finite $\mu_c$~\cite{Raj,Hal}. Possible
signatures of this  point in heavy--ion collisions were discussed
recently in Ref.~\cite{Ste}. The problem is, however, that the matter
produced in central heavy--ion collisions at very high energies has
rather low net baryon density. If such a baryon--free matter would
expand following (locally) an equilibrium path, it would miss the first
order transition line. In this case no clear signatures of the phase
transition would be observed.

Moreover, it is most likely that the matter evolution in
ultrarelativistic heavy--ion collisions does not follow thermodynamical
equilibrium. The reason is that the matter produced in such collisions
expands very fast. As has been already observed in heavy--ion
experiments at the SPS energies (see e.g.~\cite{Mun}), the expansion
velocities along the beam direction are close to the speed of light and
the transverse velocities are close to $0.5\,c$\,. A strong collective
expansion of matter is expected also at RHIC and LHC energies. Under
such conditions one may expect significant deviations from
thermodynamical equilibrium \cite{Mis1}, especially from the chemical
equilibration. This may happen on both the partonic and the hadronic
stages of the reaction.

One can mention at least two mechanisms which may lead to the deviation
from chemical equilibrium on the partonic stage. First, it is believed
that multiple color strings are produced initially in hard
nucleon--nucleon collisions. Later on they decay via the Schwinger
mechanism into quark--antiquark pairs whose abundances are determined
by quark masses and a string tension constant. At this stage,
multiplicities of secondary quarks and antiquarks may be different from
their values in thermodynamical equilibrium. Second, simple fits of QCD
lattice data \cite{Hei,Car2} indicate that gluons may acquire a large
effective mass around the deconfinement transition point. Therefore,
even if the ideal QGP were created at some intermediate stage of the
reaction, gluons would subsequently decay into lighter quark--antiquark
pairs. Hence, the abundances of different quark species may deviate
strongly from their equilibrium values. In particular, an
overpopulation of the light quark--antiquark pairs may be expected. The
phase diagram of such a chemically nonequilibrated matter may be very
different from the predictions based on the equilibrium concepts.

Another motivation to study chemically nonequilibrated quark--antiquark
systems comes from the hadronic spectroscopy. It is well known that
some mesonic resonances do not fall into the classification scheme
based on the constituent quark model. They cannot be interpreted as
conventional \qq bound states. Rather they could be either bag--like
$(qq\bar{q}\bar{q})$ states or meson--meson bound states
$(q\bar{q}-q\bar{q})$. Well known examples include the $f_0(980)$
($K\bar{K}$ bound state close to the threshold), the $f_1(1420)$
($K\bar{K}^{\star}$), the $f_0(1500)$ and $f_2(1565)$ ($\omega\omega$
and $\rho\rho$) etc. \cite{PDG}.  A legitimate question is: what will
happen if more and more \qq pairs will be put together? Such multi-\qq
systems might be even more bound due to the reduced surface energy as
compared with the bulk one. The mesonic substructure will most likely
melt away and such states will look like multi-\qq bags. We call this
hypothetical state of matter as ``meso-matter'' and its finite
droplets at ``mesoballs''. The analogous state of hadronic matter,
bound multipion droplets, has been considered in Ref.~\cite{Mis93a}.

Since the direct application of QCD at moderate temperatures and
nonzero chemical potentials is not possible at present, more simple
effective models respecting some basic symmetry properties of QCD are
commonly used. One of the most popular models of this kind, which is
dealing with constituent quarks and respects chiral symmetry, is the
Nambu--Jona-Lasinio (NJL) model \cite{Nam61,Lar61}. In recent years
this model has been  widely used for describing hadron properties (see
reviews \cite{Vog91,Kle92}), phase transitions in dense matter
\cite{Asa89,Kli90a,Cug96,Kli97,Raj,Kle99} and multiparticle bound
states \cite{Koc87,Bub96,Mis96,Bub99}.

In the previous paper \cite{Mis99} we have used the NJL model to study
properties of the quark--antiquark plasma out of chemical
equilibrium. In fact, we considered a system with independent
densities of quarks and antiquarks. To our surprise, we have found not
only first order transitions but also deep bound states even in the
baryon--free matter with equal densities of quarks and antiquarks. The
consideration in Ref.~\cite{Mis99} was limited to systems composed of
either light ($u, d$) or strange ($s$) quarks and antiquarks. In
the present paper we extend the model to arbitrary mixtures of light
and strange quarks. The emphasis is put on investigating the
possibility of bound states in such systems at various flavour
compositions of quarks and antiquarks. As a special case we consider
the baryon--rich quark matter in chemical equilibrium, in particular,
the possibility of bound states in strange quark matter, i.e.
strangelets. Thermal properties of meso-matter and strange quark matter
are also studied.

The paper is organized as follows: in Sect.~II a generalized NJL model
including flavour--mixing terms is formulated in the mean--field
approximation. Then in Sect.~III the model is used to study  the bound
states in \qq systems with different strangeness contents. The model
predictions for strangelets are discussed in Sect.~IV. The
characteristics of bound states at zero temperature are summarized in
Sect.~V. Effects of finite temperatures are considered in Sect.~VI.
Possible decay modes of new bound states are discussed in Sect.~VII.
Main results of the present paper are summarized in Sect.~VIII.

\section{Formulation of the model}

We proceed from the SU(3)--flavour version of the NJL model suggested in
Ref.~\cite{Reh96}. The corresponding Lagrangian is written as
($\hbar=c=1$)
\begin{eqnarray}
{\cal L}=\psib\,(i\,\dd-\hat{m}_0)\,\psi&+&G_S\sum_{j=0}^{8}
\left[\left(\psib\,\,\frac{\ds\lambda_j}{\ds 2}\,\psi\right)^2+
\left(\psib\,\frac{\ds i\gamma_5\lambda_j}
{\ds 2}\,\psi\right)^2\right]\nonumber\\
&-&G_V\sum_{j=0}^{8}\left[
\left(\mbox{$\psib\,\gamma_\mu \frac{\ds\lambda_j}{\ds 2}\,\psi$}\right)^2+
\left(\mbox{$\psib\,\gamma_\mu \frac{\ds\gamma_5\lambda_j}
{\ds 2}\,\psi$}\right)^2\right]\nonumber\\
&-&K\left[{\rm det}_f\,\mbox{$\left(\psib\,(1-\gamma_5)\,\psi\right)$}+
{\rm det}_f\,\mbox{$\left(\psib\,(1+\gamma_5)\,\psi\right)$}\right].
\label{lagr}
\end{eqnarray}
Here~$\psi$~is~the~column~vector~consisting~of~three~single--flavour
spinors $\psi_f$, \mbox{$f=u,d,s$}, $\lambda_1,\ldots,\lambda_8$~~are
the SU(3) Gell-Mann matrices in flavour
space,~~$\lambda_0\equiv\sqrt{2/3}\,\bm{I}$, and
\mbox{$\hat{m}_0={\rm diag}(m_{0u},\,m_{0d},\,m_{0s})$}
is the matrix of bare (current) quark masses. At $\hat{m}_0=0$ this
Lagrangian is invariant with respect to
\mbox{${\rm SU_{\,L}(3)}\otimes\,{\rm SU_{\,R}(3)}$}
chiral transformations. The second and third terms in \re{lagr}
correspond, respectively, to the scalar--pseudoscalar and
vector--axial-vector 4--fermion interactions. The last 6--fermion
interaction term breaks the $U_A(1)$ symmetry and gives rise  to the
flavour mixing effects. In particular, this term is responsible for the
large $\eta^{\prime}$ mass \cite{Kli90b}.

In the mean--field approximation the Lagrangian (\ref{lagr}) is reduced
to
\begin{eqnarray}
{\cal L}_{\rm mfa}&=&\sum_f\psib_f\,(i\dD-m_f)\,\psi_f\nonumber\\
&-&\frac{\ds G_S}{\ds 2}\sum_f\rho_{\mbs{Sf}}^2 +
\frac{\ds G_V}{\ds 2}\sum_f{\rho_{\mbs{Vf}}}^2 +
4K\prod_f\rho_{\mbs{Sf}}\,,\label{lagrm}
\end{eqnarray}
where $\dD=\dd+i\,\gamma_0 G_V\rho_{\mbs{Vf}}$ and
\begin{eqnarray}
\rho_{\mbs{Sf}}&=&<\psib_f\psi_f>\,,\label{densc}\\
\rho_{\mbs{Vf}}&=&<\psib_f\gamma_0\psi_f>\label{densv}
\end{eqnarray}
are scalar and vector densities of quarks with flavour $f$\,. Angular
brackets correspond to the quantum--statistical averaging. The
constituent quark masses, $m_f$, are determined by the coupled set of
gap equations
\bel{gape}
m_f=m_{0f}-G_S\,\rho_{\mbs{Sf}}+
2K\,\prod_{f'\neq f}\rho_{\mbs{Sf'}}\,.
\ee

The NJL model is non-renormalizable, because its coupling constants
have non-trivial dimensions: $G_S,~G_V\propto [{\rm mass}]^{-2}$ and
$K\propto [{\rm mass}]^{-5}$. As a consequence, the contribution of
negative energy states of the Dirac sea are divergent, and one must
introduce an ultraviolet  cut--off. In this respect the NJL model is an
effective model, aimed at describing the non-perturbative regime of QCD
at low energies. Following common practice, we introduce the
3--momentum cut--off $\Lambda$ to regularize divergent integrals. The
structure of the fermionic vacuum within the NJL model is shown
schematically in Fig.~1(a). Only ``active'' levels of the Dirac sea,
i.e. with $p<\Lambda$ are included in calculations.

The model parameters $m_{0f}, G_S, K, \Lambda$ can be fixed by
reproducing the observed masses of $\pi, K$\,, and $\eta'$ mesons as
well as the pion decay constant $f_\pi$. As shown in Ref.~\cite{Reh96},
a reasonable fit is achieved with the following values:
\begin{eqnarray}
m_{0u}=m_{0d}&=&5.5~{\rm MeV},~~~m_{0s}=140.7~{\rm MeV},\\
G_S=20.23~{\rm GeV}^{-2}, &&K=155.9~{\rm GeV}^{-5},
~~~\Lambda=0.6023~{\rm GeV}.\label{papa}
\end{eqnarray}
Motivated by the discussions in in Refs.~\cite{Vog91,Cas95},
we choose the following ``standard'' value of the vector coupling
constant{\footnote{This value is somewhat different from the one used
in our previous paper~\cite{Mis99}, although the ratio $G_V/G_S$
is the same.}}
\bel{coup}
G_V=0.5\,G_S=10.12\,{\rm GeV}^{-2}\,.
\ee
It should be stressed that the vector and axial--vector terms cannot
simply be ignored in the effective Lagrangian~(\ref{lagr}), as it is
often done. These terms are necessary for correctly describing the
vector meson properties \cite{Kli90b,Pol95}, for adjusting the nucleon
axial charge $g_A$~\cite{Lut92}, etc. In context of the present study,
the vector interaction is important, because it generates a net
repulsive contribution in asymmetric matter, i.e. when the numbers of
quarks and antiquarks are not equal.

Let us consider homogeneous, thermally (but not, in general,
chemically) equilibrated quark--antiquark matter at temperature $T$\,.
Let $a_{\bm{p},\lambda}$\,(\,$b_{\bm{p},\lambda}$) and
$a^+_{\bm{p},\lambda}$\,(\,$b^+_{\bm{p},\lambda}$) be the destruction
and creation operators of a quark (an antiquark) in the state
$\bm{p},\lambda$\,, where $\bm{p}$ is the 3-momentum and $\lambda$ is
the discrete quantum number denoting spin and flavour (color indices
are suppressed). By using the plane wave decomposition of quark spinors
in~\re{lagrm}, it can be shown~\cite{Mis99} that quark and antiquark
phase--space occupation numbers coincide with the Fermi--Dirac
distribution functions:
\begin{eqnarray}
<a^+_{\bm{p},\lambda}\,a_{\bm{p},\lambda}>\equiv\ocnf
&=&\left[\,\exp{\left(\frac{E_{\bm{p}f}-\mu_{Rf}}
{T}\right)}+1\,\right]^{-1}\,,\label{ocnq}\\
<b^+_{\bm{p},\lambda}\,b_{\bm{p},\lambda,f}>\equiv\ocnbf &=&
\left[\,\exp{\left(\frac{E_{\bm{p}f}-\mub_{Rf}}{T}\right)}
+1\,\right]^{-1}\,,\label{ocnqb}
\end{eqnarray}
where
$E_{\bm{p}f}=\sqrt{m_f^2+\bm{p}^2}$ and $\mu_{Rf},\,\mub_{Rf}$
denote the reduced chemical potentials of quarks and antiquarks:
\begin{eqnarray}
\mu_{Rf}&=&\mu_f-G_V\rho_{\mbs{Vf}}\,,\label{rce1}\\
\mub_{Rf}&=&\mub_f+G_V\rho_{\mbs{Vf}}\,.\label{rce2}
\end{eqnarray}
In our calculations we consider the chemical potentials $\mu_f$ and
$\mub_f$ as independent variables. The assumption of chemical
equilibrium with respect to creation and annihilation of \qq pairs
would lead to the conditions
\bel{eqcp}
\mub_f=-\mu_f,~~f=u,d,s~.
\ee

The explicit expression for the vector density can be written as
\bel{vecd}
\rho_{Vf}=\rho_f-\rhob_f\,,
\ee
where
\bel{denf}
\rho_f=\nu\intp\,\ocnf,\,\,\,\,\rhob_f=\nu\intp\,\ocnbf\,
\ee
are, respectively, the number densities of quarks and antiquarks
of flavour $f$ and $\nu=2 N_c=6$ is the spin--color degeneracy factor.
The net baryon density is obviously defined as
\bel{bard}
\rho_B=\frac{1}{3}\sum_f\rho_{Vf}\,.
\ee
The physical vacuum ($\rho_f=\rhob_f=0$) corresponds
to the limit $\ocnf=\ocnbf=0$\,.

Within the NJL model the energy density and pressure of matter as
well as the quark condensates $\rho_{Sf}$ contain divergent terms
originating from the negative energy levels of the Dirac sea. As noted
above, these terms are regularized by introducing the 3--momentum cutoff
\mbox{$\theta\,(\Lambda -|\bm{p}|)$}, where
$\theta\,(x)\equiv\frac{1}{2}\,(1+{\rm sgn}\,x)$.  Then the scalar
density is expressed as
\bel{cdens}
\rho_{Sf}=\nu\intp\frac{m_f}{E_{\bm{p}f}}\left[\,\ocnf+\ocnbf-
\theta\,(\Lambda-p)\,\right]\,.
\ee

The energy density and pressure are obtained in a standard way
from the energy-momentum tensor corresponding to the Lagrangian
(\ref{lagrm}). They can be decomposed into several parts as
\begin{eqnarray}
e&=&e_K+e_D+e_S+e_V+e_{FM}+e_0\,,\label{enden}\\
P&=&P_K+P_D+P_S+P_V+P_{FM}+P_0\,,\label{prest}
\end{eqnarray}
These expressions include:\\
the ``kinetic'' terms
\begin{eqnarray}
e_K&=&\nu\sum_f\intp\,E_{\bm{p}f}\left(\ocnf+
\ocnbf\right)\,,\label{ekin}\\
P_K&=&\frac{\nu}{3}\sum_f\intp\frac{\bm{p}^2}
{E_{\bm{p}f}}\,\left(\ocnf+\ocnbf\right)\,,\label{pkin}
\end{eqnarray}
the ``Dirac sea'' terms
\bel{pdir}
e_D=-P_D=-\nu\sum_f\intp\,E_{\bm{p}f}\,\theta\,(\Lambda -p),
\ee
the scalar interaction terms
\bel{pscal}
e_S=-P_S=\frac{\ds G_S}{\ds 2}\sum_f\rho_{\mbs{Sf}}^{\,2}\,,
\ee
the vector interaction terms
\bel{pvect}
e_V=P_V=\frac{\ds G_V}{\ds 2}\sum_f\rho_{\mbs{Vf}}^{\,2}
\ee
and the flavour mixing terms
\bel{pfmix}
e_{FM}=-P_{FM}=-4K\prod_f\rho_{\mbs{Sf}}\,.
\ee
A constant $e_0=-P_0$ is introduced in Eqs. (\ref{enden}) and
(\ref{prest}) in order to set the energy density and pressure of the
physical vacuum equal to zero. This constant can be expressed through
the vacuum values of constituent masses, $m_f^{\rm vac}$, and quark
condensates, $\rho_{\mbs{Sf}}^{\rm vac}$. These values are obtained by
selfconsistently solving the gap equations (\ref{gape}) in vacuum, i.e.
at $\ocnf=\ocnbf=0$\,.

For a system with independent chemical potentials for quarks ($\mu_f$) and
antiquarks ($\mub_f$) one can use the thermodynamic identity
\bel{tiden}
e=\sum_f(\mu_f\rho_f+\mub_f\rhob_f)-P+sT\,.
\ee
Then one can obtain the standard expression for the entropy density,
\bel{entr}
s=\partial_{\,T} P_K=
-\nu\,\sum_f\intp\left[\,\ocnf\ln{\ocnf}+(1-\ocnf)\ln{(1-\ocnf)}
+ \ocnf\to\ocnbf\right]\,.
\ee
By using Eqs.~(\ref{gape}), (\ref{ocnq})--(\ref{entr}) one can also show
that the differential relation
\bel{dpre}
dP=\sum_f\left(\rho_f\,d\mu_f+\rhob_f\,d\mub_f\right) +s\,dT
\ee
holds for any thermally (but not necessarily chemically) equilibrated process.

\section{Symmetric quark--antiquark matter}

In this section we study the multi--quark--antiquark systems at zero
temperature. Let us consider a symmetric system with equal numbers
of quarks and antiquarks for each flavour. This requirement enforces the
chemical potentials of quarks and antiquarks to be equal,
\bel{chemp}
\mub_f=\mu_f~.
\ee
In this case the net vector density is automatically zero for each
flavour, $\rho_{Vf}=0$. The net baryon number and electric charge are
also zero. Such systems are especially interesting because the
contribution of repulsive vector interaction, Eq.~(\ref{pvect}),
vanishes in this case. One can view such systems as a compressed meson
gas where mesons are  melted to their elementary constituents, quarks
and antiquarks. The single particle states of quarks and antiquarks in
such a system are schematically shown in Fig.~1(b). To understand this
picture one should simply realize that antiquarks are holes in the
Dirac sea. Then one can imagine that a certain number of quarks from
the negative energy states are collectively excited into the positive
energy states, producing an equal number of holes. Due to the presence
of valence quarks and antiquarks the mass gap will be reduced. In this
situation one can expect the appearance of bound states.

To characterize the flavour composition we introduce the strangeness
fraction parameter
\bel{rs}
r_s=\frac{N_s+N_{\bar{s}}}{N_u+N_{\bar{u}}+N_d+N_{\bar{d}}+N_s+
N_{\bar{s}}}\,,
\ee
where $N_{f(\bar{f})}$ is the number of quarks (antiquarks) of flavour
$f$\,. For simplicity we consider only the isospin--symmetric
mixtures where $N_u=N_d$ and $N_{\bar{u}}=N_{\bar{d}}$.

Fig.~2 shows the energy per particle, $\epsilon=e/\rho_{\rm tot}$, as a
function of total density of valence quarks and antiquarks,
\mbox{$\rho_{\rm tot}=\sum\limits_f\,(\rho_f+\rhob_f)$,} for different
$r_s$.  At low densities $\epsilon$ tends to the sum of the constituent
quark and antiquark masses in vacuum, weighted according to $r_s$,
\bel{vmrs}
\epsilon (\rho_{\rm tot}\to 0, r_s)=m_q^{\rm vac}(r_s)=
(1-r_s)\,m_u^{\rm vac}+r_s\,m_s^{\rm vac}.
\ee
With growing  density, $\epsilon$ first decreases due to the attractive
scalar interaction. At higher densities $\epsilon$ starts to increase,
approaching slowly the limit of ideal ultrarelativistic Fermi gas.  For
each $r_s$ one can see the appearance of a nontrivial minimum
corresponding to a bound multiparticle state with
$\epsilon_{\rm min}\,(r_s)<m_q^{\rm vac}(r_s)$. The density of \qq
pairs in these bound states varies between 0.7 and 1.4 fm$^{-3}$,
depending on $r_s$. It is easy to see~\cite{Mis99} that the minimum in
$\epsilon$ at any $r_s$ corresponds to zero pressure. Thus, finite
droplets of such matter could be in mechanical equilibrium with vacuum.

A more detailed behaviour of $\epsilon$ in the plane
$\rho_{u+d}-\rho_s$ is shown in Fig.~3(a)
(here \mbox{$\rho_{u+d}=\rho_u+\rho_d=2\,\rho_u$}). The condition
of fixed $r_s$ corresponds to a straight line with slope $r_s /(1-r_s)$
starting from the origin. By inspecting the figure one can notice the
valley of local {minima\footnote{This valley is shown by the dotted
line in Fig.~3(a).}} starting  from $\epsilon=0.482$ GeV per particle
in the pure $s\bar{s}$ system ($r_s=1$) and descending to
$\epsilon=0.304$ GeV per particle in pure $u\bar{u}+d\bar{d}$ matter
($r_s=0$). There is no potential barrier on the way from $r_s=1$ to
$r_s=0$. Thus, a droplet with any $r_s\neq 0$ will eventually ``roll
down'' in the state with $r_s=0$. Possible decay modes of the
considered bound states are discussed in Sect.  VII.

The constituent masses of $u$ and $s$ quarks as functions of
$\rho_{tot}$ are shown in Fig.~4 for different $r_s$. As expected, the
constituent masses decrease with density, signaling the gradual
restoration of chiral symmetry. It is interesting to note that the
constituent mass $m_f$ is more sensitive to the density of the same
flavour $f$. For instance, in pure $u\bar{u}+d\bar{d}$ matter, i.e. at
$r_s=0$, the $s$-quark mass does not change much with density. In this
case $m_s$ varies entirely due to the flavour--mixing interaction. At
densities corresponding to the bound states (indicated by dots on the
respective curves) the constituent masses drop significantly as
compared to their vacuum values:
$m_{u,d}\simeq 0.1\,m_{u,d}^{\rm vac}$\,,\,$m_s\simeq 0.3\,m_s^{\rm vac}$.
For zero bare masses, $m_{0f}=0$, the constituent quark masses in bound
states would be practically zero. This means that the bound states
correspond to the Wigner phase where chiral symmetry is restored.

Fig.~5 shows the chiral condensate $\langle\bar{u}u\rangle$ which, in
our notation (see Eq.~(\ref{densc})), coincides with the scalar
density $\rho_{Su}$ (notice the minus sign on the vertical axis). Its
behaviour is strongly correlated  with the $u$--quark constituent mass
(compare with Fig.~4). As before, the variation of the
$\langle\bar{u}u\rangle$ condensate is mainly sensitive to the density
of valence $u$--quarks and antiquarks. At $r_s=1$, when this density is
zero, the change in the $\bar{u}u$ condensate is caused by the
flavour--mixing interaction. The behaviour of the
$\langle\bar{s}s\rangle$ condensate is qualitatively similar, but due
to the larger bare mass, $m_{0s}\simeq 140$ MeV, its density dependence
is weaker.

The $u$ and $s$ chemical potentials are shown in Fig.~6. It is
interesting that initially they drop and then rise with density. One
can easily understand this behaviour by analyzing the explicit
expression for $\mu_f$,
\bel{muf}
 \mu_f=\sqrt{m_f^{\,2}+p_{Ff}^{\,2}}~,
\ee
where $p_{Ff}=\left(6\pi^2\rho_f/\nu\right)^{1/3}$ is the Fermi
momentum of quarks with flavour $f$ (antiquarks have the same Fermi
momentum in symmetric matter). At low densities, when $p_{Ff}\ll m_f$
and $\mu_f\simeq m_f$, all chemical potentials decrease with density
together with the respective constituent mass $m_f$ (see Fig.~4). At
high densities, when $p_{Ff}\gg m_f$, the chemical potential grows with
density as in the free relativistic Fermi--gas,
$\mu_f\simeq p_{Ff}\propto\rho_f^{1/3}$.  In fact, this nontrivial
behaviour of $\mu_f$ is responsible for the first order phase
transition discussed below.

\section{Strange quark matter}

Let us turn now to quark matter with nonzero net baryon density at
$T=0$. We assume that only valence quarks are present, i.e. the density
of valence antiquarks is zero for each flavour ($\rhob_{f}=0$).
In other words, this  means that all levels in the Dirac sea are filled
up, and additionally some levels in the Fermi sea are occupied
by the valence quarks. This situation is schematically shown in
Fig.~1(c). In contrast to the symmetric \qq matter discussed above, now
the symmetry between positive and negative energy states is lost due to
the presence of repulsive vector interaction.

Fig.~7 shows the energy per baryon $\epsilon$ as a function of baryon
density, $\rho_B=\frac{1}{3}\sum\limits_f\rho_f$\,. Different curves
correspond to different $r_s$, which in this case is simply the ratio
of the strange quark density to the total density of all quarks. At
$\rho_B\to 0$ the energy per quark tends to the corresponding vacuum
mass  given by Eq.~(\ref{vmrs}). With growing density both the
attractive scalar and repulsive vector interactions contribute to
$\epsilon$. It is interesting that at $r_s\leq 0.7$ the attractive
interaction is strong enough to produce a nontrivial local minimum at a
finite $\rho_B$. In the pure $u, d$ matter ($r_s=0$) this minimum is
unbound by about 20 MeV as compared  to the vacuum masses of $u$ and
$d$ quarks. On the other hand, it is located at a baryon density of
about $1.8\,\rho_0$, which is surprisingly close to the  saturation
density of normal nuclear matter. Of course, the location of this
minimum depends on the model parameters. Nevertheless, one can
speculate that nucleon--like 3--quark correlations, not considered in
the mean--field approach, will turn this state into the correct nuclear
ground state.

When $r_s$ grows from 0 to about 0.4, the local minimum is getting
more pronounced and the corresponding baryon density increases to about
$3.2\,\rho_0$. At larger $r_s$, the minimum again becomes more shallow
and disappears completely at $r_s\simeq 0.7$. At \mbox{$0.2<r_s<0.6$}
the minima correspond to the true bound states, i.e. the energy per
quark is lower than the respective vacuum mass. But in all cases these
bound states are rather shallow: even the most strongly  bound state at
$r_s\simeq 0.4$ is bound only by about 5 MeV per quark or 15~MeV per
baryon. Nevertheless, the appearance of local minima signifies the
possibility for finite droplets to be in mechanical equilibrium with
the vacuum at $P=0$. It is natural to identify such droplets with
strangelets, which are hypothetical objects made of light and strange
quarks \cite{Wit84,Far84,Gre87,Gil93,Schaf98,Wilczek}. Similar
multiparticle bound states made of nucleons and hyperons were also
discussed \cite{Schaf90,Schaf93,Schaf94}.

It should be emphasized here that $\beta$--equilibrium is not
required in the  present approach (see the discussion below).
That is why our most bound strangelets are predicted to be more
rich in strange quarks ($r_s>1/3$) than in the approaches assuming
$\beta$--equilibrium \cite{Wit84,Far84,Gil93}, which give
$r_s<1/3$. As a result, these strangelets will be negatively
{charged\footnote{Negatively--charged strangelets have been also
considered in Refs.~\cite{Gre87,Schaf98}.}}. Indeed, the ratio of
the charge $Q$ to the baryon number $B$ is expressed through $r_s$
as
\bel{q}
\frac{Q}{B}=\frac{2}{3}\frac{\rho_u}{\rho_B}-\frac{1}{3}
\frac{\rho_d}{\rho_B}
-\frac{1}{3}\frac{\rho_s}{\rho_B}=\frac{1}{2}(1-3r_s).
\ee
For $r_s\simeq 0.4$ this gives $Q/B\simeq -0.1$. In light of recent
discussions (see e.g.~Ref.~\cite{Wilczek}) concerning possible
dangerous scenarios of the negatively--charged strangelet production at
RHIC, we should emphasize that the strangelets predicted here are not
absolutely {bound\footnote{The analogous conclusion has been made in
Ref.~\cite{Bub99}.}}, i.e. their energy per baryon is higher than that
for the normal nuclear matter. Hence, the spontaneous conversion
of normal nuclear matter to strange quark matter is energetically
not possible.

The behaviour of the energy per quark for arbitrary mixtures of light
and strange quarks is shown in Fig.~3(b). One can clearly see that the
valley corresponding to the local minima has always a positive slope in
$r_s$. There is no potential barrier separating the states with
$r_s\neq 0$ and $r_s=0$. Thus, a strangelet with any $r_s\neq 0$ will
freely roll down to the normal nuclear matter state with $r_s=0$. It is
interesting to note that a barrier in $r_s$ may appear if attractive
scalar interaction is arbitrarily enhanced in the strange
sector~\cite{Schaf98}.

Fig.~8 shows the constituent masses of $u$ and $s$ quarks as functions
of baryon density. The dropping masses manifest again a clear tendency
to the restoration of chiral symmetry at high densities. As before,
the dots indicate the masses at the local minima in the respective
energies per baryon shown in Fig.~7. These mass values are somewhat
larger than in the symmetric \qq matter discussed above (see Fig.~4).
Note that the stronger is reduction of constituent masses the deeper
are the corresponding bound states. For the metastable state at
$r_s=0$, which is a candidate for the nuclear ground state, the masses
of $u$ and $s$ quarks are equal to 0.3 and 0.9 of their vacuum values
respectively. For the most bound state at $r_s\simeq 0.4$ the
respective mass ratios are reduced to 0.15 and 0.6.

The behaviour of the $u$ and $s$ chemical potentials is shown in
Fig.~9. One can notice the differences as compared with symmetric \qq
matter (Fig.~6). In particular, the curves for different $r_s$ do not
intersect. Due to the additional contribution of the vector
interaction, which is  linear in $\rho_B$ (see Eq.~(\ref{rce1}) where
$\mu_{Rf}=\sqrt{m_f^{\,2}+p_{Ff}^{\,2}}$ at $T=0$), the minima in
$\mu_u$ and $\mu_s$ are less pronounced. At $r_s>0.7$ the curves have
no minima at all. Therefore the chiral phase transition will be not
as strong in this case as in symmetric matter.

\section{Systematics of bound states}

The properties of the multiparticle bound states discussed in
preceding sections are summarized in Figs.~10--11. Fig.~10 shows the
binding energy per quark,
$m_q^{\rm vac}(r_s) -\epsilon_{\rm min}(r_s)$,
where $m_q^{\rm vac}$ is the energy at $\rho_{tot}=0$ and
$\epsilon_{\rm min}(r_s)$ is the energy at a local  minimum, both taken
for a given $r_s$ value. To avoid misunderstanding, Figs.~11(a) and
11(b) show $\epsilon_{\rm min}(r_s)$ separately.  Although the absolute
minimum of $\epsilon_{\rm min}$ corresponds to $r_s=0$, the maximum
bindings are realized at nonzero $r_s$. For symmetric $\bar{q}q$
systems the maximum value of about 90 MeV is reached for $r_s\simeq 0.7$.
This is indeed very strange and very bound matter! For asymmetric
systems, where $\rhob_f=0$, the maximum binding energy is much
smaller, about 5~MeV per quark at $r_s\simeq 0.4$. One should bear in
mind, however, that in this case $\epsilon_{\rm min}$ results from a
strong cancellation between the attractive scalar and repulsive vector
interactions. Therefore, it is very sensitive to their relative
strengths. The results presented above are obtained for $G_V=0.5\,G_S$.
For comparison in Figs.~10 and 11(b) we also present the model
predictions for $G_V=0$. In this case the maximum binding energy
increases to about 30~MeV per quark and the corresponding $r_s$ value
shifts to about 0.6. It is interesting to note that for
\mbox{$G_V=0$} the bound state appears even in the pure $u,d$ matter.
The corresponding binding energy is about 7 MeV per quark, i.e. 21 MeV
per baryon.

Figures~10 and 11 reveal some differences compared to the previous
calculations in Ref.~\cite{Mis99}. This is mainly due to a different
set of model parameters used there. In particular, the present set of
parameters gives higher quark constituent masses in the vacuum,
\mbox{$m_{u,d}^{\rm vac}\simeq 368$ MeV} and
$m_s^{\rm vac}\simeq 550$ MeV, as compared to $m_{u,d}^{\rm vac}=300$
MeV and $m_s^{\rm vac}=520$~MeV in Ref.~\cite{Mis99}. This leads to an
overall upward shift in the energy per particle by 30--60~MeV. On the
other hand, inclusion of the flavour--mixing interaction lowers the
energy. As a result, all binding energies increase slightly. A local
minimum in $\epsilon$ appears now even in the pure $u, d$ matter.

The dots in Fig.~11 indicate the positions of some conventional mesons
(a) and baryons~(b). Their empirical masses are rescaled (by factor
$1/2$ for mesons and $1/3$ for baryons) and shown at $r_s$ values
corresponding to their flavour compositions. By inspecting the figure,
one can make a few interesting observations. According to Fig.~11(a),
mesoballs lie lower in energy than conventional vector mesons, but
higher than the pseudoscalar mesons (see discussion in Sect.~VIII). As
one can see in Fig.~11(b), conventional baryons are more bound than
strangelets even at $G_V=0$\,. This is an indication that baryon--like
3--quark correlations might be indeed very important in the
baryon--rich quark matter.

\section{Quark matter at finite temperatures}

In this section we study properties of deconfined matter at finite
temperatures. The calculations for this case can be done by using
general formulas of Sect.~II with the quark and antiquark occupation
numbers given by Eqs.~(\ref{ocnq})--(\ref{ocnqb}).

First we discuss thermal properties of meso-matter where the chemical
potentials obey the conditions (\ref{chemp}). A typical behaviour of
the equation of state for this case is illustrated in Fig.~12(a). It
shows the pressure isotherms for symmetric $q\bar{q}$ matter with
strangeness content $r_s=1/3$.  One can clearly see a strong first
order phase transition which is signaled by the appearance of isotherms
with negative slopes, $\partial_\rho P<0$ (spinodal instability). The
corresponding critical temperature is about 100 MeV.  Another important
feature is that the zero-pressure states persist up to temperatures as
high as 70 MeV. This means that a finite droplet of this matter which
has cooled down to this temperature and still remaining at high density
of $q\bar{q}$ pairs (around 4\,$\rho_0$ in this case), will be trapped
in a bound state. At later times it will further cool down by emitting
hadrons from the surface (see below).

The critical temperatures for a phase transition and for the appearance
of a bound state are shown in Fig.~13(a) as functions of $r_s$. One can
conclude that this dependence is rather weak: the first critical
temperature changes between 90 and 110 MeV while the second one varies
around 70 MeV.

Now let us consider chemically--equilibrated quark matter at finite
temperatures. In this case the chemical potentials of quarks obey the
conditions (\ref{eqcp}). Fig.~12(b) represents the pressure isotherms
for the case $\mu_s=0$\,. This condition implies equal numbers of
strange quarks and antiquarks, i.e. zero net strangeness. It is
appropriate for fast processes where strangeness is conserved, e.g. in
relativistic nuclear collisions. As before one can see a region of
spinodal instability, \mbox{$\partial_\rho P<0$}, which is
characteristic for a first order phase transition. However, this phase
transition is much weaker than in the case of symmetric $q\bar{q}$
matter (note the different scales in Figs.~12(a) and 12(b)). The
corresponding critical temperature is about 35 MeV in this case. The
zero--pressure states exist only at temperatures below 15~MeV.

Generally, the equation of state of the chemically equilibrated quark
matter is characterized by two quantities: net baryon charge, $B$, and
net strangeness, $S$. Therefore, it is interesting to study thermal
properties of this matter at $S\neq 0$\,. Such states can be
reached in neutron stars. They can also be realized via the
distillation mechanism accompanying a QCD phase transition in heavy--ion
collisions \cite{Car91}. Leaving a detailed study of this question for
a future publication, here we only present results which complement our
discussion concerning Fig.~13(a). Fig.~13(b) shows the critical
temperatures for the phase transition and for the bound states in the
equilibrated matter as functions of $r_s^-=S/3B$\,\footnote{We
introduce this new notation in order to distinguish this quantity from
$r_s$ defined in Eq.~(\ref{rs}). In chemically equilibrated matter at
$T=0$ (no antiquarks) these two quantities coincide.}. One can see that
both temperatures first grow with $r_s^-$ and then drop to zero at
$r_s^-\sim 0.8$. The maximal values of respectively 50 MeV and 30 MeV
are realized at some intermediate $r_s^-$ around 0.4. As demonstrated
earlier in Fig.~10, this value of $r_s^-$ corresponds to the most bound
state of strange quark matter at $T=0$. So we see an obvious
correlation: the deeper is a bound state at $T=0$ the stronger is a
phase transition at finite temperatures.

It should be emphasized here again that the thermal properties of
asymmetric baryon--rich quark matter are very sensitive to the relative
strength of scalar and vector interactions. To illustrate this point we
have calculated the phase diagrams for several values of~$G_V$.  The
results for $\mu_s=0$ are presented in Fig.~14. It shows the boundaries
of two--phase coexistence regions calculated by using standard Gibbs
conditions~\cite{Mis99}. If we take $G_V=0$, as in most calculations
in the literature, the coexistence region becomes wider and the
corresponding critical temperature increases to about 70 MeV. On the
other hand, if one takes $G_V=0.65\,G_S$, the phase transition becomes
weaker: the critical temperature moves down to 20 MeV and the
zero--pressure states disappear completely. The calculation shows that
there is no phase transition at $G_V>0.71\,G_S$\,. It is interesting to
note that in all cases this first order phase transition occupies the
region of densities around the normal nuclear density $\rho_0$\,.  For
instance, at $G_V=0.5\,G_S$ the coexistence region at $T=0$ extends
from 0.2~to 1.7~$\rho_0$.

These results demonstrate that this chiral phase transition is rather
similar to a liquid--gas phase transition in normal nuclear matter.
The critical temperature and baryon density in the present case
($T_c\sim 30$ MeV, $\rho_{Bc}\sim\rho_0$) are not so far from the
values predicted by the conventional nuclear models~\cite{Goo85}
($T_c\sim 20$ MeV, $\rho_{Bc}\sim 0.5\,\rho_0$)\,. One may expect that
in a more realistic approach, taking into account nucleonic
correlations, the ``chiral'' transition may turn into the ordinary
``liquid--gas'' phase transition. If this would be the case, one
should be doubtful about the possibility of any other QCD phase
transition of the liquid--gas type at a higher baryon density. At least
only one phase transition of this type is predicted within the NJL
model.

\section{Discussion of decay modes}

Let us discuss briefly possible decay channels of the novel states of
quark matter described above. Naively one could think that the
symmetric \qq matter would be extremely unstable with respect to the
annihilation of quarks and antiquarks. But, in fact, many annihilation
channels are closed or do not exist in dense \qq matter.  From the
first sight, one may think that two--pion annihilation,
\mbox{$\bar{q}+q\to\pi+\pi$}, should be the strongest decay channel in
the bulk (one--pion annihilation is not allowed by the
energy--momentum conservation). But simple arguments show that this
might be not true. Indeed, the bound states of \qq matter appear at
such high densities when chiral symmetry is practically restored. In
this case the pion looses its special nature as a Goldstone boson and
inside this matter it should be as heavy as other mesons. Moreover, due
to the Mott transition \cite{Zhu94}, it is unlikely that conventional
mesonic states survive in this dense medium. Therefore, one can expect
that the hadronic channels of the \qq annihilation simply do not exist
in the bulk.

Another possible decay channel is the strong flavour conversion,
\mbox{$\bar{s}s\to\bar{u}u$} or $\bar{d}d$. In principle, this process
can proceed through the one--gluon (color octet) intermediate state.
However, it may be shown that due to color neutrality the corresponding
matrix elements vanish after summation over the color indices. Within
the present approach, the 4--fermion interaction terms are diagonal in
the flavour space (see Eq.~(\ref{lagrm})) and, therefore, cannot change
the flavour. Strong flavour conversion is possible only through the
6--fermion flavour--mixing interaction,
\mbox{$\bar{s}s\to\bar{u}u+\bar{d}d$}\,.  This channel is only open
when $\mu_s>\mu_u+\mu_d$. By inspecting Fig.~6 one can see that this
condition is never fulfilled. Therefore we conclude that strong flavour
conversion is not possible or at least strongly suppressed in the bulk.

In a finite droplet, hadronic annihilation,
$\bar{q_{f}}+q_{f'}\to h_1+h_2+...$, is certainly  possible at the
surface, if $\bar{\mu}_{f}+\mu_{f'}>m_{h_1}+m_{h_2}+...$\,. It is clear
from Fig.~11(a) that the emission of pions in annihilation of light \qq
pairs is energetically more favorable than the emission of heavier
mesons. Moreover, many annihilation channels of strange quarks and
antiquarks are simply closed e.g.
\mbox{$\overline{s}+s\to K+\overline{K}$}\,.
As a result of the pion emission, the strangeness content of the
daughter droplet will increase\footnote{Similar processes leading to
strangeness distillation have been considered in
Refs.~\cite{Gre87,Car91}.}. Finally, it will contain predominantly the
$s\bar{s}$ pairs, similarly to a system composed of $\phi$--mesons
(``$\phi$--ball'').

Further hadronization may be slowed down by several reasons
\cite{Mis99}. According  to Fig.~10, at $r_s \simeq 1$ the energy
available in the annihilation of a $s\bar{s}$ pair is
$\mu_s+\bar{\mu}_s\simeq 0.95 $ GeV. This is lower than the threshold
energies of the hadronic states $\phi(1020)$ and $K\bar{K}(990)$. The
only open channels in this case are $\rho\pi(910)$ and $3\pi(420)$.
Their partial width in the $\phi$-meson decay is about 0.7 MeV. If this
would be the dominant decay mode, one could expect a life time for
these $\phi$--balls of about 280 fm/c, i.e much longer than the typical
duration of a heavy--ion collision.

In strange quark matter (without antiquarks), flavour conversion is
only possible through weak decays. As follows from Fig.~9, at densities
corresponding to zero pressure, the condition \mbox{$\mu_s>\mu_u$}
holds. This means that weak processes of the types
\mbox{$s\to u+e^{-}+\overline{\nu}_e$\,,}\,\,\mbox{$s+u\to u+d$} are
allowed. Since there is no local barrier at any $r_s$ (see
Fig.~11(b)), a system produced initially at some $r_s\neq 0$ will
subsequently reduce this $r_s$ value by a cascade of weak decays. As a
result, the system will roll down along the line of local minima shown
in Fig.~11(b) (see also Fig.~3(b)). Finally, all $s$ quarks will
be converted into light $u, d$ quarks. Two steps of this conversion
process are shown schematically in Figs.~15(a) and~15(b).

This picture is very different as compared to the one based on the
MIT bag model. Within that model the pressure and energy density are
given by  simple expressions
\bel{mit}
P_{\rm MIT}=P_{\rm id}-B,~~~e_{\rm MIT}=e_{\rm id}+B~,
\ee
where $B$ is a bag constant, $P_{\rm id}$ and $e_{\rm id}$ are the
pressure and energy density for a mixture of ideal gases of $u$, $d$
and $s$ quarks with constant (bare) masses,
\mbox{$m_{u, d}=m_{0u,d}\simeq 0$}, \mbox{$m_s=m_{0s}\simeq 150$ MeV}.
By properly choosing $B$ one can always get a zero pressure point and,
accordingly, a minimum in the energy per baryon. Because the quark
masses are kept constant, the condition $\mu_s>m_s$ will be first
satisfied at a relatively low baryon density $\propto m_{0s}^3$. At
higher densities a certain fraction of $s$ quarks will always be
present in a $\beta$--equilibrated matter.

In contrast, in the NJL model the $s$ quark mass is a function of both
baryon density and strangeness content. As one can see from Fig.~9, at
any given $r_s$ the condition $\mu_{u, d}=\mu_s$ can be satisfied only
at sufficiently high baryon densities which correspond to positive
pressure. On the other hand, at the points of zero pressure we always
have $\mu_{u, d}<\mu_s$. Therefore, weak decays will proceed until a
strangelet reaches $r_s=0$ (see Fig.~15(c)).

\section{Conclusions}

In the present paper we have investigated properties of deconfined
matter at different densities of quarks and antiquarks, not necessarily
constrained by the conditions of chemical equilibrium.  All
calculations are carried out within the SU(3)--flavour NJL model
including scalar, vector and flavour--mixing interactions. We have
demonstrated the possibility of strongly bound states in symmetric \qq
systems consisting of equal numbers of quarks and antiquarks
(``mesoballs'').  The maximal binding energies of mesoballs, of about
90 MeV per particle or 180 MeV per \qq pair, are realized for flavour
compositions with about 70\% of $s\overline{s}$ pairs. These systems
remain bound up to the temperatures $T\loo 70$\,MeV. The lifetimes of
mesoballs may be long enough due to the suppression  of annihilation
into hadrons in the bulk. The model predicts a strong first order phase
transition in chemically nonequilibrated meso-matter, i.e. at zero net
baryon density. The critical temperature of this phase transition is
in the range of 90--110 MeV depending of the relative abundance of
$s\overline{s}$ pairs.  As discussed in Ref.~\cite{Mis99}, formation of
mesoballs in high--energy heavy--ion collisions may be observed through
the event--by--event analysis of $\pi, K, \phi$ spectra. In particular,
narrow bumps in the hadron rapidity distributions may be generated by
hadronizing mesoballs.

By using the same model, we have also investigated the equation of
state of chemically equilibrated deconfined matter at various
temperatures, baryon densities and strangeness contents. The model
predicts the existence of loosely bound, negatively--charged
strangelets with maximal binding energies of about 20 MeV per
baryon at $r_s\sim 0.4$\,. Similarly to Ref.~\cite{Bub99}, no
absolutely stable strange quark matter has been found. It is shown that
properties of baryon--rich quark matter are very sensitive to the
relative magnitude of the vector and scalar interactions. At the
standard value of vector and scalar couplings, $G_V/G_S=0.5$, the
metastable bound states of chemically equilibrated matter exist at
$T<15$ MeV, while at $G_V=0$ this temperature increases to 40 MeV.

The first order chiral phase transition in equilibrated baryon--rich
quark matter is much weaker as compared to the symmetric $q\bar{q}$
matter. In the case of zero net strangeness the critical temperature
is in the range of 30 MeV and the critical baryon density is
around~$\rho_0$. We believe that this phase transition is reminiscent
of the ordinary liquid--gas phase transition in nuclear matter.
One can use the present model to study the chiral phase transition at
nonzero net strangeness, which is appropriate e.g. for neutron star
matter. We have found that the maximal critical temperature
$T_c\simeq 50$ MeV is reached for the ratio of net strangeness to
baryon charge $S/B\simeq 1.2$\,.

It should be emphasized that all calculations in this paper have been
made with constant values of coupling constants $G_S, G_V, K$\,.  We
think that, this is a reasonable approximation at intermediate
densities and temperatures. On the other hand, the asymptotic freedom of
QCD requires all effective interactions to vanish at high densities and
temperatures. Therefore, in a more realistic approach the coupling
constants should decrease with density and temperature. Hopefully, this
behaviour can be simulated by introducing finite range form factors.

\section*{Acknowledgments}

The authors thank M. Belkacem and J. Schaffner-Bielich for useful
discussions. Two of us (I.N.M. and L.M.S.) thank the Institut
f\"ur Theoretische Physik, J.W.~Goethe~Universit\"at, Frankfurt am
Main, for the kind hospitality. This work has been  supported
by the Alexander von Humboldt Stiftung, the Graduiertenkolleg
``Experimentelle und Theoretische Schwerionenphysik'',
GSI, BMBF and DFG.

\section *{Figure captions}
\newcommand{\Fig}[2]{\noindent{FIG.~#1.~}
\parbox[t]{15cm}{\baselineskip 24pt #2}\\[5mm]}
\Fig{1}
{Schematic picture of energy levels of quarks and antiquarks in vacuum
(a), symmetric \qq droplet (b) and pure quark droplet (c). Occupied
quark states are shown by full dots while open dots represent holes
(antiquarks). $R$ denotes the radius of a droplet, $m$ and $V$ are,
respectively, constituent mass and vector potential of quarks. The
boundaries of vacuum mass gap are shown by dashed lines.}
\Fig{2}
{Energy per particle $\epsilon$ in symmetric \qq matter vs. total
particle density at different values of strangeness fraction $r_s$
(shown near the corresponding curves). Points indicate local minima of
$\epsilon$.  $\rho_0=0.17\,{\rm fm}^{-3}$ is normal nuclear density.}
\Fig{3}
{Contours of energy per particle $\epsilon$ (in GeV) for symmetric (a)
and asymmetric (b) matter in the $\rho_{u+d}-\rho_s$ plane. Dotted
lines represent local extrema of $\epsilon$\,. Shading shows regions
with negative pressure.}
\Fig{4}
{Constituent masses of $u$ and $s$ quarks in symmetric \qq matter as
functions of total particle density. Figures in the box show values of
strangeness fraction $r_s$. Dots correspond to minima of energy per
particle at given $r_s$.}
\Fig{5}
{The same as in Fig. 4, but for condensate densities of $u$ and $s$ quarks.}
\Fig{6}
{The same as in Fig. 4, but for chemical potentials of $u$ and $s$ quarks.}
\Fig{7}
{The same as in Fig. 2, but for asymmetric $u, d, s$ matter without
antiquarks.}
\Fig{8}
{The same as in Fig. 4, but for asymmetric quark matter.}
\Fig{9}
{The same as in Fig. 6, but for asymmetric quark matter.}
\Fig{10}
{Binding energies per particle in symmetric \qq matter (solid line) and
asymmetric quark matter as functions of strangeness fraction $r_s$.
Dotted line shows the results of calculations when the vector
interaction is switched off ($G_V=0$).}
\Fig{11}
{Minimal energies per particle in symmetric (a) and asymmetric (b)
matter as functions of strangeness fraction $r_s$. Dotted lines show
energy per particle in the limit of zero particle densities,
Eq.~(\ref{vmrs}). Different parts of the solid line in the lower panel
correspond to metastable (AB and CD) or bound (BC) states.
Triangles and squares show masses of lightest mesons and
baryons devided by the total number of constituents (two in mesons
and three in baryons). Symbol $<K>$ represents the spin averaged mass
of $K$ and $K^*(892)$ mesons, i.e.  $(3m_{K^*}+m_K)/8$.  Dashed line in
the lower plot shows the results in the limit $G_V\to 0$.}
\Fig{12}
{Pressure isotherms for symmetric \qq matter (a) and equilibrated
quark matter with \mbox{$\mu_s=0$}~(b). Temperatures are given in MeV
near the corresponding curves. Boundaries of spinodal regions are shown
by the dashed lines.}
\Fig{13}
{Critical temperatures for existence of bound states ($P<0$) and
phase transition ($\partial_\rho P<0$) in symmetric \qq
matter (a) and equilibrium quark matter (b) as functions
of strangeness fraction.}
\Fig{14}
{Phase diagrams of equilibrium quark matter with zero net strangeness
at various ratios of vector and scalar coupling constants.}
\Fig{15}
{Schematic pictures of energy levels (shown by shading) occupied by
light and strange quarks in cold quark matter at different strangeness
fractions $r_s$. Hatched boxes represent the constituent quark masses.
Arrows show weak decay processes \mbox{$s\to u+e+\overline{\nu}_e$} and
\mbox{$s+u\to u+d$}.}
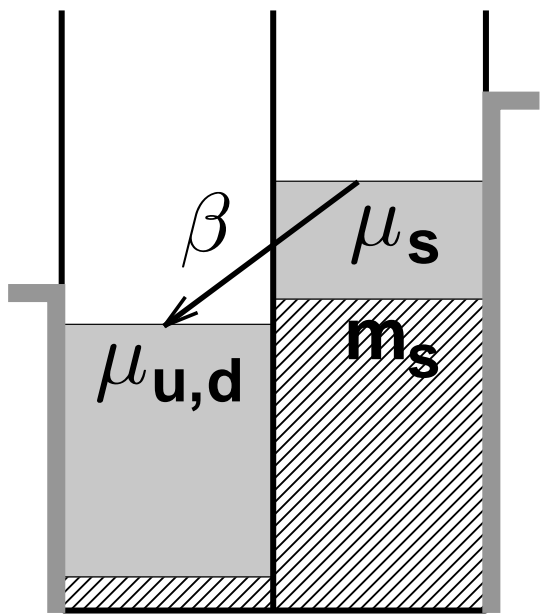
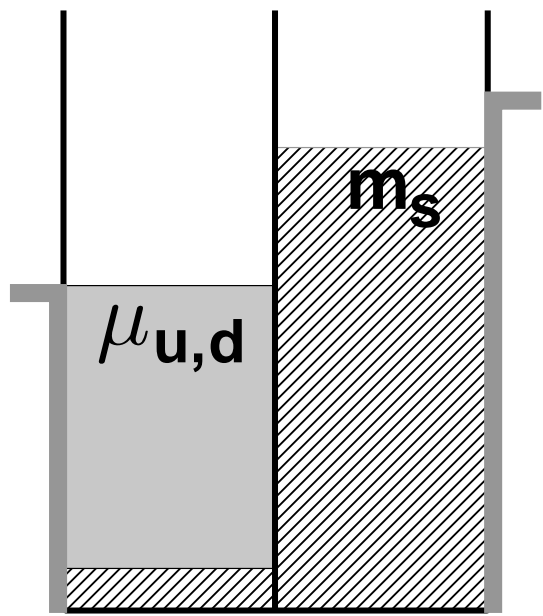
\end{document}